\documentclass{PoS}
\usepackage{amsmath}
\usepackage{amssymb}
\usepackage{mcite}
\newcommand{\path}[1]{ #1}

\newcommand{\psibar}[0]{\overline{\psi}}
\newcommand{\phibar}[0]{\overline{\phi}}

\newcommand{\Jbar}[0]{\overline{J}}
\newcommand{\Bbar}[0]{\overline{B}}
\newcommand{\Bbarm}[0]{\overline{B}^{\;-1}}
\newcommand{\Bm}[0]{{B}^{-1}}
\newcommand{\Bbareta}[0]{\overline{B}^{(\eta)}}
\newcommand{\Bbaretam}[0]{(\overline{B}^{(\eta)})^{-1}}
\newcommand{\Betam}[0]{(B^{(\eta)})^{-1}}
\newcommand{\Betaa}[0]{{B}^{(\eta)}}

\newcommand{\cbar}[0]{\overline{c}}
\newcommand{\Tr}[0]{\text{Tr}}
\newcommand{\sign}[0]{\text{sign}}
\newcommand{\ket}[1]{\left|#1\right\rangle}
\newcommand{\bra}[1]{\left\langle#1\right|}
\newcommand{\braket}[2]{\left\langle #1 | #2 \right\rangle}
\title{A Ginsparg-Wilson approach to lattice $\mathcal{CP}$ symmetry}

\ShortTitle{lattice $\mathcal{CP}$ symmetry}

\author{\speaker{Nigel Cundy}\\
        Institut f\"ur Theoretische Physik, Universit\"at Regensburg, D-93040 Regensburg, Germany\\ \\
        Lattice Gauge Theory Research Center, FPRD, and CTP, Department of Physics \&
    Astronomy, Seoul National University, Seoul, 151-747, South Korea\\ \\
        E-mail: \email{ndcundy@phya.snu.ac.kr}}

\abstract{There is a long standing challenge in lattice QCD concerning the relationship between $\mathcal{CP}$-symmetry and lattice chiral symmetry: na\"ively the chiral symmetry transformations are not invariant under $\mathcal{CP}$. With results similar to a recent work by Igarashi and Pawlowski, I show that this is because charge conjugation symmetry has been incorrectly realised on the lattice. The naive approach, to directly use the continuum charge conjugation relations on the lattice, fails because the renormalisation group blockings required to construct a doubler free lattice theory from the continuum are not invariant under charge conjugation. Correctly taking into account the transformation of these blockings leads to a modified lattice $\mathcal{CP}$ symmetry for the fermion fields, which, for gauge field configurations with trivial topology, has a smooth limit to continuum $\mathcal{CP}$ as the lattice spacing tends to zero. After constructing $\mathcal{CP}$ transformations for one particular group of lattice chiral symmetries, I construct a lattice chiral gauge theory which is $\mathcal{CP}$ invariant and whose measure is invariant under gauge transformations and $\mathcal{CP}$.}

\FullConference{The XXVIII International Symposium on Lattice Field Theory, Lattice2010\\
		June 14-19, 2010\\
		Villasimius, Italy}

\begin{document}

\section{Introduction}
The problem of simulating chiral symmetry on the lattice has been solved using the Ginsparg-Wilson procedure to derive an alternative lattice chiral symmetry. The symmetry is defined for a lattice Dirac operator $D$ by a Ginsparg-Wilson equation~\cite{Ginsparg:1982bj},
\begin{gather}
\gamma_L D + D\gamma_R = 0,
\end{gather}
where $\gamma_L$ and $\gamma_R$ are local operators which reduce to $\gamma_5$ in the continuum limit and which satisfy $\gamma_L^2 = \gamma_R^2 = 1$, and a corresponding lattice chiral symmetry~\cite{Luscher:1998pqa}
\begin{align}
\psibar \rightarrow& \psibar e^{i\epsilon \gamma_L}&\phantom{space} \psi = e^{i\epsilon\gamma_R}\psi.
\end{align}
The simplest known practical lattice Dirac operator whose action is invariant under a lattice chiral symmetry is the overlap operator~\cite{Narayanan:1993sk,*Narayanan:1993ss,*Neuberger:1998fp},
\begin{gather} 
D = 1+ \gamma_5\sign(K), 
\end{gather}
for some suitable kernel operator $K$. Conventionally, the symmetry is expressed using the choice $\gamma_L = \gamma_5$ and $\gamma_R = \gamma_5 (1-D)$. In practice, however, there is an infinite group of (non-commuting) chiral symmetries satisfied by any Ginsparg-Wilson Dirac operator, including the overlap operator~\cite{Mandula:2007jt,*Mandula:2009yd}.

A conceptual difficulty with these chiral symmetry relations is that they are not obviously $\mathcal{CP}$-invariant. $\mathcal{CP}$ symmetry has been applied to the lattice in its continuum form, which converts 
\begin{align}
\mathcal{CP}:\psibar\rightarrow& \psi^TW&
\mathcal{CP}:\psi\rightarrow& -W^{-1}\psibar^T\nonumber\\
\mathcal{CP}:\gamma_5 \rightarrow& - W^{-1}\gamma_5^TW&
\mathcal{CP}: D[U,x,y] \rightarrow& W^{-1}D[U^{\mathcal{CP}},\overline{x},\overline{y}]^TW.\label{eq:contcp}
\end{align}
For the lattice action to simultaneously respect chiral symmetry and $\mathcal{CP}$ symmetry would require $\mathcal{CP}:\gamma_R \rightarrow - \gamma_L$ and $\mathcal{CP}:\gamma_L \rightarrow - \gamma_R$, which has been shown to be impossible for any local $\gamma_L$ and $\gamma_R$ with the correct continuum limit~\cite{Fujikawa:2002is,*Jahn:2002kg}. The lattice Dirac operator obeys the same $\mathcal{CP}$ transformation law as the continuum operator~\cite{Cundy:2010pu}.

In these proceedings, I suggest that this anomaly is caused because $\mathcal{CP}$ symmetry (or more specifically charge conjugation symmetry) has been incorrectly realised on the lattice, and, the correct form of the symmetry follows naturally from the same Ginsparg-Wilson procedure used to construct lattice chiral symmetry. In section \ref{sec:2}, I review the Ginsparg-Wilson renormalisation group procedure as used to construct overlap fermions; in section \ref{sec:3} I use these results to write down the modified lattice $\mathcal{CP}$ relations in the trivial topological sector, and, in section \ref{sec:4}, I construct a chiral gauge theory. I conclude and give an outlook in section \ref{sec:5}.
\section{Renormalisation group blockings}\label{sec:2}
Working exclusively in the continuum, for a generating functional, with Dirac operator $D_0$
\begin{gather}
Z = \int d\psi_0 d\psibar_0 dU e^{-S_g[U] + \psibar_0 D_0 \psi_0 + \Jbar_0 \psi_0 + \psibar_0 J_0},
\end{gather}
where $S_g$ is some representation of the Yang-Mills action, we can transform to a new fermionic field $\psi_1$ with Dirac operator $D$ using invertible blocking matrices $B$, $\Bbar$ and $\alpha$,
\begin{align}
Z =& \int dU \frac{1}{\det{\alpha}}\int d\psi_0 d\psibar_0 e^{-S_g[U] - \psibar_0 D_0 \psi_0 + \Jbar_0 \psi_0 + \psibar_0 J_0}\int d\psi_1 d\psibar_1 e^{-(\psibar_1 - \psibar_0\Bbarm) \alpha(\psi_1 - \Bm\psi_0)}\nonumber\\
\propto & \int d\psi_1 d\psibar_1 dU e^{-S'_g[U] - \psibar_1 D \psi_1 + \Jbar_1 \psi_1 + \psibar_1 J_1},
\end{align}
with
\begin{gather}
D = \alpha - \alpha {\Bm}\frac{1}{D_0 + {\Bbar\alpha {B}}}\Bbarm\alpha.
\end{gather}
As $\alpha \rightarrow \infty$, this reduces to
\begin{gather}
D = \Bbar D_0 B.
\end{gather} 
If the original action is invariant under chiral symmetry,
\begin{align}
\psibar_0 \rightarrow& \psibar_0 e^{i\epsilon \gamma_5}&\psi_0\rightarrow e^{i\epsilon \gamma_5} \psi_0,
\end{align}
then by expanding in infinitesimal $\epsilon$ it is straightforward to show that the new Dirac operator obeys
\begin{gather}
\Bbar\gamma_5 \Bbarm D + D \Bm \gamma_5 B = D (\alpha^{-1}\Bbar\gamma_5 \Bbarm D + \Bm \gamma_5 B \alpha^{-1} )D. \label{eq:2.6}
\end{gather} 
Using $\alpha \rightarrow \infty$ (so that the right hand side of (\ref{eq:2.6}) vanishes), $\Bbar^{(\eta)} = D^{(1-\eta)/2} Z D_0^{(-(1-\eta)/2}$ and $B^{(\eta)} =D_0^{(-(1+\eta)/2}Z^{-1}D^{(1+\eta)/2}$, with $Z$ defined as a function which commutes with $\gamma_5$ and maps the eigenvectors of $D_0$ onto the eigenvectors of $D$~\cite{Cundy:2009ab}, one obtains the family of chiral symmetries,
\begin{align}
\gamma_L^{(\eta)} = &\Bbareta\gamma_5 \Bbaretam & \gamma_R^{(\eta)} = &\Betam \gamma_5 \Betaa. 
\end{align}
This formulation only works, of course, if the blocking matrices $B$ and $B^{-1}$ are invertible, which requires (although this condition is not sufficient) that the blocking is between two Dirac operators with the same rank, i.e. a lattice theory to another lattice theory with (possibly) a different lattice spacing and physical volume but the same lattice size, or a continuum theory to another continuum theory. To block from the continuum to a lattice requires that we express the lattice Dirac operator as the smooth limit of an equivalent continuum Dirac operator (for example, with the same dispersion relation and the same renormalisation constants). This can be achieved by decomposing the Dirac operator into lattice and off-lattice components (e.g. one could use Schur's procedure, although my own work uses a different and more convenient decomposition), and then giving the off-lattice section an infinite mass in the continuum limit so that it leaves all physics unaffected and can be safely neglected in a numerical simulation. This procedure will, of course, not work for any arbitrary lattice Dirac operator --- even if the Dirac operators are of the same rank, that does not necessarily or usually imply that the blockings are finite and invertible, but it does work for overlap fermions~\cite{Cundy:2009ab}, as strongly implied by the observation that overlap fermions do, in fact, satisfy the Ginsparg-Wilson equation with local $\gamma_L$ and $\gamma_R$.

In the topological sector with no zero modes\footnote{The method outlined in this work struggles in other topological sectors because of difficulties with relating the lattice doublers associated with the zero modes with eigenvalues of the continuum Dirac operator. This causes a non-locality in the lattice $\mathcal{CP}$ transformations (while a non-local lattice $\mathcal{CP}$ symmetry is tolerable, the  lattice $\mathcal{CP}$ needs to correctly reduce to continuum $\mathcal{CP}$ and here the non-locality causes a problem). The issue is that the lattice operator, which describes two fermion fields, one physical and one doubler, is mapped to a single physical continuum fermion. In future work, I will discuss how this approach is modified when a `doubler' fermion is added to the continuum action~\cite{Cundy:2010pu,cundyforthcoming}.},  the overlap and continuum Dirac operators can be decomposed into a basis defined by the non-zero chiral eigenvectors of $D^{\dagger} D$
\begin{align}
D =& \sum_i\lambda_i\left(\begin{array}{l l} \ket{g^+_i}&\ket{g^-_i}\end{array}\right)
\left(\begin{array}{l l}
\cos\theta& \sin\theta\\
-\sin\theta& \cos\theta
\end{array}\right)\left(\begin{array}{l } \bra{g^+_i}\\\bra{g^-_i}\end{array}\right) \nonumber\\
%%%
D_0 =&\sum_i \lambda_{0,i}\left(\begin{array}{l l} \ket{\tilde{g}^+_i}&\ket{\tilde{g}^-_i}\end{array}\right)
\left(\begin{array}{l l}
\cos\pi/2& \sin\pi/2\\
-\sin\pi/2& \cos\pi/2
\end{array}\right)\left(\begin{array}{l } \bra{\tilde{g}^+_i}\\\bra{\tilde{g}^-_i}\end{array}\right), \label{eq:matrixdecomp}
\end{align} 
where  \begin{align}
Z =& \sum_i\ket{g^+_i}\bra{\tilde{g}^+_i} + \ket{g^-_i}\bra{\tilde{g}^-_i},&
\tan \theta =& \frac{2\sqrt{1-D^{\dagger}D/4}}{\sqrt{D^{\dagger} D}},
\end{align}
and $\ket{g^{\pm}}$ are defined (up to a constant phase fixed by equation (\ref{eq:matrixdecomp})) by the equations
\begin{align}
D^{\dagger}D \ket{g^{\pm}_i} =& \lambda_i^2\ket{g^{\pm}_i}, & \gamma_5\ket{g^{\pm}_i} =& \pm \ket{g^{\pm}_i},&\braket{g^{\pm}}{g^{\pm}} =& 1,\nonumber\\ 
D_0^{\dagger}D_0 \ket{\tilde{g}^{\pm}_i} =& \lambda_{0,i}^2\ket{\tilde{g}^{\pm}_i}, & \gamma_5\ket{\tilde{g}^{\pm}_i} =& \pm \ket{\tilde{g}^{\pm}_i},&\braket{g^{\pm}}{g^{\mp}} =& 0.
\end{align}
It is now straight-forward to construct a practical form for $\gamma_L$ and $\gamma_R$~\cite{Cundy:2010pu},
\begin{align}
\gamma_R^{(\eta)} =& \gamma_5 \cos((\eta+1)(\theta- \pi/2)) + \sign(\gamma_5(D^{\dagger}-D))\sin((\eta+1)(\theta- \pi/2))\nonumber\\
\gamma_L^{(\eta)} =& \gamma_5 \cos((\eta-1)(\theta- \pi/2)) + \sign(\gamma_5(D^{\dagger}-D))\sin((\eta-1)(\theta- \pi/2)).
\end{align}
These operators are local only when $(\eta+1)/2$ is an integer~\cite{Cundy:2010pu} (at the zero eigenvalues of $\gamma_5(D^{\dagger} - D)$, $\theta = 0$ or $\pi/2$, so $\sin((\eta+1)(\theta- \pi/2)) = 0$ only for these values of $\eta$). The conventional  Ginsparg-Wilson relation is the solution at $\eta = 1$. It can be shown that these operators have the correct continuum limit, are Hermitian, unitary, and satisfy the Ginsparg-Wilson relation. This group of lattice chiral symmetries is that discussed by Mandula~\cite{Mandula:2007jt,*Mandula:2009yd}.
\section{Application to $\mathcal{CP}$ symmetry}\label{sec:3}
The transformations of the blockings under $\mathcal{CP}$ symmetry follow directly from the known transformations of the Dirac operators and $\gamma_5$:
\begin{align}
\mathcal{CP}:B^{(\eta)} \rightarrow& W^{-1}(\Bbar^{(-\eta)})^T W,&\mathcal{CP}:\Bbar^{(\eta)} \rightarrow& W^{-1} (B^{(-\eta)})^T W,\\
\mathcal{CP}:\gamma_R^{(\eta)} \rightarrow & -W^{-1} (\gamma_L^{(-\eta)})^TW,&\mathcal{CP}:\gamma_L^{(\eta)} \rightarrow & -W^{-1} (\gamma_R^{(-\eta)})^TW.
\end{align}
The blocked fermion field is given by
\begin{align}
\psi_1^{(\eta)} =& (B^{(\eta)}) ^{-1} \psi_0,& \psibar_1^{(\eta)} = & \psibar_0 (\Bbar^{(\eta)}) ^{-1},
\end{align}
and thus
\begin{align}
\mathcal{CP}:\psi_1^{(\eta)} \rightarrow& W^{-1}(\psibar_1^{(\eta)} \Bbar^{(\eta)}(\Bbar^{(-\eta)})^{-1})^T,&\mathcal{CP}:\psibar_1^{(\eta)} \rightarrow& -((B^{(-\eta)})^{-1} B^{(\eta)}\psi_1^{(\eta)})^TW.
\end{align}
By writing $B$ and $\Bbar$ in terms of their matrix decomposition, following the method outlined in equation (\ref{eq:matrixdecomp}), one obtains
\begin{align}
\Bbar^{(\eta)}(\Bbar^{(-\eta)})^{-1} =& \gamma_R^{(\eta - 1)}\gamma_5,&((B^{(-\eta)})^{-1} B^{(\eta)} = \gamma_5\gamma_R^{(\eta - 1)},
\end{align}
which gives
\begin{align}
\mathcal{CP}:\psi_1^{(\eta)} \rightarrow&- W^{-1}(\psibar_1^{(\eta)} \gamma_R^{(\eta-1)}\gamma_5)^T,&\mathcal{CP}:\psibar_1^{(\eta)} \rightarrow& ( \gamma_5\gamma_R^{(\eta-1)}\psi_1^{(\eta)})^T W. 
\end{align}
It can be shown that~\cite{Cundy:2010pu}
\begin{align}
[\gamma_5\gamma_R^{(\eta-1)},D] =& 0,& \phantom{space}
\gamma_R^{(\eta-1)}\gamma_5\gamma_R^{(-\eta)} \gamma_5\gamma_R^{(\eta-1)} =& \gamma_R^{(\eta)},
\end{align}
and hence
\begin{align}
&\mathcal{CP}: \psibar^{(\eta)}_1 D (1+\gamma_R^{(\eta)})\psi^{(\eta)}_1 \rightarrow \psibar^{(\eta)}_1 \gamma_R^{(\eta-1)}\gamma_5D(1+\gamma_R^{(-\eta)})\gamma_5 \gamma_R^{(\eta-1)} \psi^{(\eta)}_1=\psibar^{(\eta)}_1 D (1+\gamma_R^{(\eta)})\psi^{(\eta)}_1,\nonumber\\
&\mathcal{CP}: \psibar^{(\eta)}_1 D \psi^{(\eta)}_1 \rightarrow \psibar^{(\eta)}_1 D \psi^{(\eta)}_1.
\end{align}
Therefore both the standard and chiral gauge Lagrangians are invariant under this lattice $\mathcal{CP}$.
\section{Weyl fermions}\label{sec:4}
To construct Weyl fermions, the measure should be invariant under both gauge transformations and $\mathcal{CP}$. On the lattice, the measure depends on the operators $\gamma_R^{(\eta)}$ and therefore the gauge field. The fermion field can be decomposed in terms of a complete basis $\phi^{(\eta,+)}$ and $\phi^{(\eta,-)}$, where
\begin{align}
\psi^{(\eta)} =& \sum_i[c^{(\eta,+)}_i \phi^{(\eta,+)}_i + \phi^{(\eta,-)}_i c^{(\eta,-)}_i],&\phantom{space}
\psibar^{(\eta)} =& \sum_i[\cbar^{(\eta,+)}_i \phibar^{(\eta,+)}_i + \phibar^{(\eta,-)}_i \cbar^{(\eta,-)}_i].\label{eq:measure}
\end{align}
The measure is then $|d\psi^{(\eta)} d\psibar^{(\eta)}| = |\prod dc_i^{(\eta,+)} dc_i^{(\eta,-)} d\cbar_i^{(\eta,+)} d\cbar_i^{(\eta,-)}|$. The measure of a single Weyl fermion is constructed from (for example) $\phi^{(\eta,+)}_i$ and $\phibar^{(\eta,-)}_i$.
It is convenient to write $\phi^{(\eta,\pm)}$ as the eigenvectors of $\gamma_R^{(\eta)}$ and $\phibar^{(\eta,\pm)}$ as the eigenvectors of $\gamma_L^{(\eta)}$, so that
\begin{align}
\gamma_R^{(\eta)}\phi^{(\eta,\pm)}_i =& \pm \phi^{(\eta,\pm)}_i, & \gamma_L^{(\eta)}\phibar^{(\eta,\pm)}_i =& \pm \phibar^{(\eta,\pm)}_i
\end{align}
Under $\mathcal{CP}$, these eigenvalue equations transform to 
\begin{align}
-W^{-1}(\gamma_L^{(-\eta)})^TW(\phi^{(\eta,\pm,\mathcal{CP})})^T = &\pm(\phi^{(\eta,\pm,\mathcal{CP})}_i)^T 
\nonumber\\
-W^{-1}(\gamma_R^{(-\eta)})^TW(\phibar^{(\eta,\pm,\mathcal{CP})})^T = &\pm(\phibar^{(\eta,\pm,\mathcal{CP})}_i)^T
\end{align}
or
\begin{align}
\mathcal{CP}:\phi^{(\eta,\pm)}_i \rightarrow&  W^{-1}(\phibar^{(-\eta,\mp)}_i)^T,&\phantom{space}
\mathcal{CP}:\phibar^{(\eta,\pm)}_i \rightarrow&  W^{-1}(\phi^{(-\eta,\mp)}_i)^T.
\end{align}
Since
\begin{gather}
\phi_i^{(-\eta,\pm)} = \gamma_5 \gamma_R^{(\eta-1)} \phi_i^{(\eta,\pm)},
\end{gather}
the transformation of the measure for a single Weyl fermion can be calculated by applying $\mathcal{CP}$ symmetry to equation (\ref{eq:measure}) :
\begin{align}
\psibar^{(\eta)}\gamma_5\gamma_R^{(\eta-1)} =& \sum_i[c^{(\eta,+,\mathcal{CP})}_i \phibar^{(-\eta,-)}_i],&\phantom{space}
\gamma_R^{(\eta-1)}\gamma_5\psi^{(\eta)} =& \sum_i[\phi^{(-\eta,}_i \cbar^{(\eta,-,\mathcal{CP})}_i],
\end{align}
and 
\begin{align}
\mathcal{CP}:c^{(\eta,+)}_i \rightarrow& \cbar^{(\eta,-)}& \mathcal{CP}:\cbar^{(\eta,-)}_i \rightarrow& c^{(\eta,+)},
\end{align}
and this measure is invariant under $\mathcal{CP}$. 
 
The change in the basis after an infinitesimal change in the basis vectors $\phi^{(\eta,+)}_j \rightarrow  \phi^{(\eta,+)}_j + \delta_{\xi}\phi^{(\eta,+)}_j$, $\phibar^{(\eta,-)}_j \rightarrow  \phibar^{(\eta,-)}_j + \delta_{\xi}\phibar^{(\eta,-)}_j$ is given by
$e^{-i\mathcal{L}}$,
where~\cite{Luscher:1998du}
\begin{gather}
\mathcal{L} = i\sum_j \left[(\phi^{(\eta,+)}_j,\delta_{\xi}\phi^{(\eta,+)}_j) + (\delta_{\xi}\phibar^{(\eta,-)}_j,\phibar^{(\eta,-)}_j)\right].
\end{gather}
If the change in the basis is caused by a change in the Dirac operator $D\rightarrow D + \delta_{\xi} D$, then, by writing $\phi$ and $\phibar$ in terms of the eigenvectors $H^+$ and $H^-$ of $\gamma_5 D$ and considering the changes of those eigenvectors under changes of the Dirac operator~\cite{Cundy:2007df}, it is possible to show that~\cite{Cundy:2010pu}
\begin{align}
(\phi^{(\eta,+)}_i,\delta_{\xi}\phi^{(\eta,+)}_i) =& \frac{\sin (2\alpha^{(\eta)})}{2}\left[(H^{-}_i,\delta_{\xi}H^+_-) + (H^{+}_i,\delta_{\xi}H^-_-)  \right]\nonumber\\
=&\frac{\sin (2\alpha^{(\eta)})}{4\lambda_i}\left[(H^{-}_i,\gamma_5\delta_{\xi}(D)H^+_-) - (H^{+}_i,\gamma_5\delta_{\xi}(\gamma_5) H^-_-)  \right]\nonumber\\
=&\frac{1}{4}\Tr \left[\gamma_5 \delta_\xi (D) \sin (2\alpha^{(\eta)}) \frac{D^{\dagger} - D}{D^{\dagger}D \sqrt{1-\frac{D^{\dagger}D}{4}}}\right],
\end{align}
with
\begin{gather}
\alpha^{(\eta)} = (\theta + (\eta + 1)(\pi/2 - \theta))/2.
\end{gather}
For an infinitesimal gauge transformation $\xi$ in a representation $R(\xi)$, the change in the Dirac operator is given by $\delta_{\xi} D = [D,R(\xi)]$, and
\begin{gather}
(\phi^{(\eta,+)}_i,\delta_{\xi}\phi^{(\eta,+)}_i) =-\frac{1}{2}\Tr \left[\gamma_5 R(\xi) \sqrt{1-\frac{D^{\dagger}D}{4}} \sin\left(2\alpha^{(\eta)}\right)\right] = 0
\end{gather}
because  $\Tr\; \gamma_5=0$  and all the other operators commute with $\gamma_5$. Therefore the measure is gauge invariant. 
\section{Conclusion}\label{sec:5}
I have proposed that, in the sector with no global topological charge, the problems associated with the realisation of $\mathcal{CP}$ symmetry with lattice chiral fermions are an illusion derived from an incorrect application of the continuum form of $\mathcal{CP}$ symmetry to the lattice. The renormalisation group blockings used to construct the lattice action are not themselves $\mathcal{CP}$ invariant, so the $\mathcal{CP}$ transformation of the lattice fermion fields must differ from the transformation of continuum fermions. Construction of a possible lattice $\mathcal{CP}$ transformation follows directly from the same Ginsparg-Wilson procedure used to establish chiral symmetry on the lattice. I have also shown that it is possible to construct a $\mathcal{CP}$ invariant chiral gauge action, where the measure is invariant under gauge transformations and $\mathcal{CP}$. Similarly a Majoranna action can be constructed, and the Higgs correctly accounted for~\cite{Cundy:2010pu}. However, the method outlined here requires modification before being extended to other topological sectors.
\section*{Acknowledgements}
I am grateful for support from the DFG SFB TR-55 and the BK21 program funded by the South Korean NRF, and for conversations with Andreas Sch\"afer and Weonjong Lee.

\bibliographystyle{elsarticle-num-mcite}
\bibliography{weyl}

\end{document}